\documentclass[11pt,a4paper]{amsart}

\usepackage{amsmath,amsfonts,amssymb}
\usepackage{mathrsfs}

\usepackage{cases}

\usepackage{svg}
\usepackage{amsthm}
\usepackage{braket}
\theoremstyle{definition}
\theoremstyle{plain}
\newtheorem*{remark}{Remark}
\newtheorem{Not}{Notation}
\newtheorem{theorem}{Theorem}[section]
\newtheorem{st}{Statement}[section]
\newtheorem*{pr}{Proof}
\newtheorem{df}{Definition}

\begin{document}

\title{Perturbations vs. Deformations}

\author{Maxim Gritskov}

\address{Krichever Center, Skolkovo Institute of Science and Technology, 121205, Moscow, Russia}
\address{Saint Petersburg State University,
Universitetskaya nab. 7/9, 199034 St. Petersburg, Russia}
\email{m.gritskov@spbu.ru}

\author{Andrey Losev}

\address{Shanghai Institute for Mathematics and Interdisciplinary Sciences, Building 3, 62
Weicheng Road, Yangpu District, Shanghai, 200433, China}
\email{aslosev2@yandex.ru}

\begin{abstract}
	In the first part of the paper we define a perturbative (pre-formal) geometry and formulate a theorem on the relation between the construction of a perturbative neighborhood of affine varieties and the higher tangent bundles. In the second part of the paper we discuss perturbative vector fields and related structures, which are finite-dimensional analogs of perturbation theory characteristics arising in quantum field theory.
\end{abstract}

\maketitle

\date{\today}

\tableofcontents

\newpage
\section{Introduction}%
\label{sec:Introduction}
The origin of this paper is quite unusual. We were studying the beta-function in perturbative functorial quantum field theory and found that what we are doing is a particular case of a pretty general mathematical construction. We could burry this construction in a long appendix to that paper, but we prefer to publish it separately. Therefore, first of all, these notes should be regarded as a detailed explanation of the procedure for constructing the perturbation theory from the deformation theory described in \cite{2}. We hope that it will appear in other cases of deformation theory.

Here we will explain the context in which we found the mathematical problem studied in this paper. Readers who are not interested
in QFT may omit this part of the paper and proceed with section 2. 

The following part of the introduction is intended for those interested in QFT origin of the problem and maybe QFT itself.
For more detail, see \cite{2}.

We work with functorial approach to QFT, where it is defined as a symmetric monoidal functor from the category of geometrically enriched cobordisms to category of linear spaces.
In simple terms, it means that we associate to a Riemann manifold with a boundary (cobordism) an element of the tensor product of linear spaces associated to boundary components. Such an element is called by physicists a partition function. Functoriality simply means that partition function behaves properly under cutting along the hypersurface not intersecting boundary components. If Riemann manifold
keeps connected after the cutting it gives linear equations on partition function, while if it decouples into two pieces it leads to quadratic equations and so on.

The universal study of all QFT would imply study of the space of general solutions to such equations, it is doable only for one-dimensional Riemann manifolds (universal description of Quantum Mechanics as $\exp(TH)$ where $T$ is the length of the interval and $H$  -  Hamiltonian operator).

For higher dimensions of Riemann manifolds, the system of equations is as universal as quadrics describing commutative associative algebras and equally useless in getting interesting examples or a kind of classification.

Keeping this in mind, physicists often start with a given QFT (say free or conformal) and deform it. One can show that the deformation corresponds to the so-called local observable of the theory. Thus, knowing the theory, we know the tangent vector to this theory in the space of theories. In other terms, we obtain the theory over the ring of dual numbers. We may iterate this process and get the theory over the ring $\mathrm{Def}_{n,k}(\epsilon)$ - see definition 3 in section 2. However, we would like to compare this iterative process with the infinitesimal neighborhood of the theory in the space of theories, i.e. theory over the ring $\mathrm{Pert}_{n,k}(\lambda)$ - see definition 1. In section 2 we show how to get the latter from the former. We do it in the language of algebraic geometry and also present explicit down-to-earth formulas for physicists who do not know that abstract nonsence.

In QFT physicists often study the so-called beta function corresponding to the change of parameters of the theory under the conformal change of the metric. It is a vector field on the space of theories, and perturbative beta-function is a vector field on the infinitesimal neighbourhood of a given theory.
Together with the $\beta$-function physicists also study $\gamma$-functions, which are operators of (anomalous) dimensions of local observables. We claim that there is a universal relation among them, and this relation follows from the relation
between two definitions of deformation theory presented in this paper. We derive this relation in some simple cases in section 3 of this paper. 

The article is organized as follows. In section 2, we define perturbative (preformal) and deformation charts of an affine variety, and formulate the main theorem connecting them. Then we deal with a down-to-earth example of constructing the perturbative neighborhood of an affine quadric using deformation charts. In section 3 we discuss the finite-dimensional analogs of the perturbative features that first appeared in quantum field theory, namely the beta and gamma functions, and their connection. 

\section{Perturbations vs. Deformations}%
\label{sec:Perturbation vs. Deformations}
\subsection{Perturbative and Deformational Charts of  Varieties}%
\label{sec:Perturbative and Deformational Charts of  Varieties}
Let $\mathbb{K}$ be $\mathbb{R}$ or $\mathbb{C}$. Consider the affine variety $X$ of dimension $n$ and $x_{\star}$ is a smooth point of the variety $X$. In addition, all morphisms must be understood in the sense of punctured affine varieties. Thus, the marked point of the spectrum of a local ring is its unique maximal ideal, and the marked point of $X$ is $x_{\star}$.
\begin{Not}
    We will always assume that the affine variety $X$ is given by the polynomial equations $F_{1}=0,...,F_{M}=0$ for some $F_{A}\in\mathbb{K}[x^{1},...,x^{N}]$. We will denote the ring of polynomial functions on variety $X$ as $P(X)$:
    \begin{equation}
        P(X)=\mathbb{K}[x^{1},...,x^{N}]/I(X).
    \end{equation}
    Here $I(X)$ is the ideal of polynomials equal to zero on $X$. In general, we can assume that this ideal is generated by $F_{1},...,F_{M}$.
\end{Not}
\begin{df}
    The ring $\mathrm{Pert}_{n,k}(\lambda)$ is defined as quotient
\begin{equation}
    \mathrm{Pert}_{n,k}(\lambda)=\mathbb{K}[\lambda^{1},...,\lambda^{n}]/(\{\lambda^{\alpha_{1}}\cdot...\cdot \lambda^{\alpha_{k+1}}; \alpha_{i}\in\overline{1,n}\}).
\end{equation}
\end{df}
\begin{df}
    A perturbative chart is an affine variety morphism
\begin{equation}
\label{pertchart}
   E:\mathrm{Spec}(\mathrm{Pert}_{n,k}(\lambda))\rightarrow X.
\end{equation}
\end{df}
\begin{remark}
If $n$ is a dimension of $X$, then the limit $k\rightarrow \infty$ of $\mathrm{Pert}_{n,k}(\lambda)$ corresponds to infinitesimal neighbourhood \cite{3} of a point $x_{\star}$ in $X$.
\end{remark}

Sometimes, it happens that it is not possible to construct a perturbative chart directly. This situation is realized in field theory, in which it is necessary to consider the intersection of an infinite number of quadrics in $\mathbb{K}^{\infty}$. However, in such a case it turns out that it is possible to construct a \emph{deformational chart}.
\begin{df}
    The ring $\mathrm{Def}_{n,k}(\epsilon)$ is defined as quotient
\begin{equation}
    \mathrm{Def}_{n,k}(\epsilon)=\mathbb{K}[\epsilon_{1}^{\alpha},...,\epsilon_{k}^{\alpha}]/(\{\epsilon_{1}^{\alpha_{1}}\cdot\epsilon_{1}^{\beta_{1}},...,\epsilon_{k}^{\alpha_{k}}\cdot\epsilon_{k}^{\beta_{k}}; \alpha_{i},\beta_{i}\in\overline{1,n}\}).
\end{equation}
\end{df}
\begin{df}
    The deformational chart is an affine variety morphism \begin{equation}
        E': \mathrm{Spec}(\mathrm{Def}_{n,k}(\epsilon))\rightarrow X.
    \end{equation}
\end{df}
\begin{remark}
    Geometrically, there is a natural one-to-one correspondence between two sets:
    \begin{equation}
        \mathrm{Hom}(\mathrm{Spec}(\mathrm{Def}_{n,k}(\epsilon)), X)\simeq (T_{x_{\star}}^{\oplus n})^{k}X=T_{x_{\star}}^{\oplus n}(...\,T_{x_{\star}}^{\oplus n}(T_{x_{\star}}^{\oplus n} X)...).
    \end{equation}
\end{remark}

It turns out that by studying all deformation charts, we will also study all perturbative ones, since for any variety $X$
\begin{equation}
    \mathrm{Hom}(\mathrm{Spec}(\mathrm{Pert}_{n,k}(\lambda)),X)\hookrightarrow \mathrm{Hom}(\mathrm{Spec}(\mathrm{Def}_{n,k}(\epsilon)),X). 
\end{equation}
In addition, we explicitly computed the image of this canonical embedding. The symmetric group $\mathrm{S}_{k}$ naturally acts on the ring $\mathrm{Def}_{n,k}(\epsilon)$ by permutations
\begin{equation}
\sigma\left(\epsilon_{i}^{\alpha}\right)=\epsilon_{\sigma(i)}^{\alpha},
\end{equation}
and this action is lifted to the action on $\mathrm{Hom}(\mathrm{Spec}(\mathrm{Def}_{n,k}(\epsilon)),X)$. We are now ready to state the theorem, the proof of which is given in appendix A.
\begin{theorem}
    There is a natural one-to-one correspondence between perturbative charts of $X$ and $\mathrm{S}_{k}$-invariant subset of $\mathrm{Hom}(\mathrm{Spec}(\mathrm{Def}_{n,k}(\epsilon)), X)$:
    \begin{equation}
        \mathrm{Hom}(\mathrm{Spec}(\mathrm{Pert}_{n,k}(\lambda)), X)\simeq\mathrm{Inv}_{\mathrm{S}_{k}}(\mathrm{Hom}(\mathrm{Spec}(\mathrm{Def}_{n,k}(\epsilon)), X)).
    \end{equation}
\end{theorem}
\noindent The corresponding theorem states that $k$-fold deformations of the exact solution $x_{\star}$ by $k$ nilpotents of index $2$ realize $k$-order perturbative solutions starting from $x_{\star}$.
\subsection{Quadric Pert- and Def- Charts}%
\label{sec:Explicit Example: Quadric}
In down-to-earth terms, the statement of the theorem of the last section says that the set of solutions of the system $F_{A}=0$ over the ring $\mathrm{Pert}_{n,k}(\lambda)$ is a subset of the $S_{k}$-invariant solutions of the same system over the ring $\mathrm{Def}_{n,k}(\epsilon)$. The corresponding embedding morphism of rings $\mathcal{I}_{\,\mathrm{S}}: \mathrm{Pert}_{n,k}(\lambda)\hookrightarrow\mathrm{Def}_{n,k}(\epsilon)$ has the form:
\begin{equation}
    \mathcal{I}_{\,\mathrm{S}}(\lambda^{\alpha})=\epsilon_{1}^{\alpha}+...+\epsilon_{k}^{\alpha}.
\end{equation}
In this section, we will demonstrate this phenomenon by a simple example. 

Let us now explain the statement of the theorem of the previous paragraph by a concrete example. Consider a single polynomial equation $F(x)=0$, where $F\in\mathbb{K}[x^{1},...,x^{N}]$. Moreover, for simplicity, assume that $F$ is a quadric.
\begin{Not}
    We introduce the following notation. Let $R$ be a ring, then by $R^{\,N}$ we denote the set of columns of elements $r^{a}\in R$, where $a\in\overline{1,N}$:
    \begin{equation}
r\in R^{\,N} \Leftrightarrow r = \left(
\begin{array}{cccc}
r^{1}\\
\vdots\\
r^{N}
\end{array}
\right).
    \end{equation}
    We also introduce the standard dot product:
    \begin{equation}
        r,s \in R^{\,N} \Rightarrow \braket{r,s}:= r^{i} s_{i}\in R.
    \end{equation}
    We will always assume a summation over repeated indices.
\end{Not}
\begin{Not}
    For $r, v_{1},v_{2}\in R^{\,N}$ we introduce the notation 
    \begin{equation}
        \partial^{2}F(r)(v_{1},v_{2})=\partial_{i_{1}}\partial_{i_{2}}F(r)\cdot v_{1}^{i_{1}}\cdot v_{2}^{i_{2}}
    \end{equation}
    is the bilinear form associated with the $2$-th derivative of $F$.
\end{Not}
Consider the problem of constructing a perturbative chart for $k=3$. The general solution of the equation $F=0$ over the ring $\mathrm{Pert}_{n,3}(\lambda)$ has the form:
\begin{eqnarray}
\label{thirdorder2}
     \notag p(\lambda)=x_{\star}+\lambda^{\alpha} s_{\alpha}-\frac{\lambda^{\alpha}\lambda^{\beta}}{2}\partial^{2}F(x_{\star})(s_{\alpha},s_{\beta})\cdot\partial F(x_{\star})- \frac{\lambda^{\alpha}\lambda^{\beta}}{2}A_{\alpha\beta}^{\gamma}\cdot e_{\gamma}+\\ +\,\frac{\lambda^{\alpha}\lambda^{\beta}\lambda^{\gamma}}{2}\partial^{2}F(x_{\star})(s_{\alpha},\partial F(x_{\star}))\partial^{2}F(x_{\star})(s_{\beta},s_{\gamma})\cdot \partial F(x_{\star})-\\ \notag-\,\frac{\lambda^{\alpha}\lambda^{\beta}\lambda^{\gamma}}{6}\left(B_{\alpha\beta\gamma}^{\delta}\cdot e_{\delta} - 3A_{\alpha\beta}^{\delta}\cdot\partial^{2}F(x_{\star})(s_{\gamma},e_{\delta})\cdot\partial F(x_{\star})\right).
\end{eqnarray}
Here $s_{\alpha}$ is the set of $n$ tangent vectors to the surface $X\subset\mathbb{K}^{N}$ at $x_{\star}$ and $A_{\alpha\beta}^{\gamma}, B_{\alpha\beta\gamma}^{\delta}$ - symbols, which are symmetric on the lower indices.
\begin{remark}
    Without loss of generality, we assume that $\braket{\partial F(x_{\star}), \partial F(x_{\star})}=1$.
\end{remark}
This result can be obtained by brute-force solution of $F=0$, or can be derived by solving some recurrence relations using the \emph{tree} diagram technique.

Let us now consider the problem of constructing a deformation chart for $k=3$. In turn, the general solution over $\mathrm{Def}_{n,3}(\epsilon)$ has the form:
\begin{eqnarray}
\label{thirdorderdef2}
p(\epsilon_{1},\epsilon_{2},\epsilon_{3})=x_{\star}+\epsilon_{1}^{\alpha}s_{\alpha}+\epsilon_{2}^{\alpha}t_{\alpha}+\epsilon_{3}^{\alpha}u_{\alpha}-\\ \notag -\,\epsilon_{1}^{\alpha}\epsilon_{2}^{\beta}\cdot\partial^{2}F(x_{\star})(s_{\alpha},t_{\beta})\cdot\partial F(x_{\star})-\\ \notag -\,\epsilon_{1}^{\alpha}\epsilon_{3}^{\beta}\cdot\partial^{2}F(x_{\star})(s_{\alpha},u_{\beta})\cdot\partial F(x_{\star})-\\ \notag -\,\epsilon_{2}^{\alpha}\epsilon_{3}^{\beta}\cdot\partial^{2}F(x_{\star})(t_{\alpha},u_{\beta})\cdot\partial F(x_{\star})-\\ \notag -\,\epsilon_{1}^{\alpha}\epsilon_{2}^{\beta}\epsilon_{3}^{\gamma}\cdot\partial^{3}F(x_{\star})(s_{\alpha},t_{\beta},u_{\gamma})\cdot\partial F(x_{\star})+\\ \notag +\,\epsilon_{1}^{\alpha}\epsilon_{2}^{\beta}\epsilon_{3}^{\gamma}\cdot\partial^{2}F(x_{\star})(s_{\alpha},t_{\beta})\partial^{2}F(x_{\star})(u_{\gamma},\partial F(x_{\star}))\cdot\partial F(x_{\star})+\\ \notag +\,\epsilon_{1}^{\alpha}\epsilon_{2}^{\beta}\epsilon_{3}^{\gamma}\cdot\partial^{2}F(x_{\star})(s_{\alpha},u_{\gamma})\partial^{2}F(x_{\star})(t_{\beta},\partial F(x_{\star}))\cdot\partial F(x_{\star})+\\ \notag +\,\epsilon_{1}^{\alpha}\epsilon_{2}^{\beta}\epsilon_{3}^{\gamma}\cdot\partial^{2}F(x_{\star})(t_{\beta},u_{\gamma})\partial^{2}F(x_{\star})(s_{\alpha},\partial F(x_{\star}))\cdot\partial F(x_{\star})+\\ \notag +\,A_{\alpha\beta}^{\gamma}\cdot\epsilon_{1}^{\alpha}\epsilon_{2}^{\beta}\epsilon_{3}^{\delta}\cdot\partial^{2}F(x_{\star})(u_{\delta},e_{\gamma})\cdot\partial F(x_{\star})-\epsilon_{1}^{\alpha}\epsilon_{2}^{\beta}\cdot A_{\alpha\beta}^{\gamma}\cdot e_{\gamma}+\\ \notag +\,B_{\alpha\beta}^{\gamma}\cdot\epsilon_{1}^{\alpha}\epsilon_{2}^{\delta}\epsilon_{3}^{\beta}\cdot\partial^{2}F(x_{\star})(t_{\delta},e_{\gamma})\cdot\partial F(x_{\star})-\epsilon_{1}^{\alpha}\epsilon_{3}^{\beta}\cdot B_{\alpha\beta}^{\gamma}\cdot e_{\gamma}+\\ \notag +\,C_{\alpha\beta}^{\gamma}\cdot\epsilon_{1}^{\delta}\epsilon_{2}^{\alpha}\epsilon_{3}^{\beta}\cdot\partial^{2}F(x_{\star})(s_{\delta},e_{\gamma})\cdot\partial F(x_{\star})-\epsilon_{2}^{\alpha}\epsilon_{3}^{\beta}\cdot C_{\alpha\beta}^{\gamma}\cdot e_{\gamma} - \\ \notag -\,\epsilon_{1}^{\alpha}\epsilon_{2}^{\beta}\epsilon_{3}^{\gamma}\cdot D^{\delta}_{\alpha\beta\gamma}\cdot e_{\delta}.
\end{eqnarray}
Here $s_{\alpha}, t_{\alpha}, u_{\alpha}\in T_{x_{\star}}X$ and it is important to note that unlike the general solution of $\mathrm{Pert}_{n,3}(\lambda)$ symbols $A_{\alpha\beta}^{\gamma}, B_{\alpha\beta}^{\gamma}, C_{\alpha\beta}^{\gamma}$ and $D_{\alpha\beta\gamma}^{\delta}$ are not symmetric on the lower indices in the general solution. This result can be obtained in the general case using the recurrence relations described in appendix B.

The ring $\mathrm{Pert}_{n,3}(\lambda)$ is a subring of $\mathrm{Def}_{n, 3}(\epsilon)$ and the embedding morphism is given by $\mathcal{I}_{\,\mathrm{S}}(\lambda^{\alpha})=\epsilon_{1}^{\alpha}+\epsilon_{2}^{\alpha}+\epsilon_{3}^{\alpha}$. This subring consists of polynomials which are invariant with respect to permutation $\epsilon_{i}^{\alpha}\mapsto \epsilon_{\sigma(i)}^{\alpha}$, where $\sigma$ is an element of the symmetric group $\mathrm{S}_{3}$. The following facts follow from the invariance of \eqref{thirdorderdef2}: $s_{\alpha}=t_{\alpha}=u_{\alpha}$, the equality of the constants $A_{\alpha\beta}^{\gamma}=B_{\alpha\beta}^{\gamma}=C_{\alpha\beta}^{\gamma}$, and furthermore, that the constants $A_{\alpha\beta}^{\gamma}$ and $D_{\alpha\beta\gamma}^{\delta}$ are completely symmetric at the lower indices. 

\begin{st}
Let $U_{\alpha\beta}$ and $V_{\alpha\beta\gamma}$ be symmetric in lower indices. Then
\begin{equation}
(\epsilon_{1}^{\alpha}\epsilon_{2}^{\beta}+\epsilon_{1}^{\alpha}\epsilon_{3}^{\beta}+\epsilon_{2}^{\alpha}\epsilon_{3}^{\beta})\cdot U_{\alpha\beta}=U_{\alpha\beta}\cdot \frac{(\epsilon_{1}^{\alpha}+\epsilon_{2}^{\alpha}+\epsilon_{3}^{\alpha})(\epsilon_{1}^{\beta}+\epsilon_{2}^{\beta}+\epsilon_{3}^{\beta})}{2}
\end{equation}
and the second identity is true:
\begin{equation}
\epsilon_{1}^{\alpha}\epsilon_{2}^{\beta}\epsilon_{3}^{\gamma}\cdot V_{\alpha\beta\gamma}=V_{\alpha\beta\gamma}\cdot\frac{(\epsilon_{1}^{\alpha}+\epsilon_{2}^{\alpha}+\epsilon_{3}^{\alpha})(\epsilon_{1}^{\beta}+\epsilon_{2}^{\beta}+\epsilon_{3}^{\beta})(\epsilon_{1}^{\gamma}+\epsilon_{2}^{\gamma}+\epsilon_{3}^{\gamma})}{6}.
\end{equation}
\end{st}
\begin{pr}
    The proof is done by direct algebraic calculations.
\end{pr}
Then denoting $\epsilon_{1}+\epsilon_{2}+\epsilon_{3}$ by $\Sigma\epsilon$ we obtain that the symmetric solutions in \eqref{thirdorderdef2} have the following form
\begin{eqnarray}
     p(\epsilon)=x_{\star}+\underbrace{(\Sigma \epsilon)^{\alpha}}_{\mathcal{I}_{\,\mathrm{S}}(\lambda^{\alpha})} s_{\alpha}-\underbrace{\frac{(\Sigma \epsilon)^{\alpha}(\Sigma \epsilon)^{\beta}}{2}}_{\mathcal{I}_{\,\mathrm{S}}(\lambda^{\alpha}\lambda^{\beta}/2)}\partial^{2}F(x_{\star})(s_{\alpha},s_{\beta})\cdot\partial F(x_{\star})-\\ \notag -\,\underbrace{\frac{(\Sigma \epsilon)^{\alpha}(\Sigma \epsilon)^{\beta}}{2}}_{\mathcal{I}_{\,\mathrm{S}}(\lambda^{\alpha}\lambda^{\beta}/2)} A^{\gamma}_{\alpha\beta}\cdot e_{\gamma} -\\ \notag -\,\underbrace{\frac{(\Sigma \epsilon)^{\alpha}(\Sigma \epsilon)^{\beta}(\Sigma \epsilon)^{\gamma}}{6}}_{\mathcal{I}_{\,\mathrm{S}}(\lambda^{\alpha}\lambda^{\beta}\lambda^{\gamma}/6)}\partial^{3}F(x_{\star})(s_{\alpha},s_{\beta},s_{\gamma})\cdot \partial F(x_{\star})+ \\ \notag +\,\underbrace{\frac{(\Sigma \epsilon)^{\alpha}(\Sigma \epsilon)^{\beta}(\Sigma \epsilon)^{\gamma}}{6}}_{\mathcal{I}_{\,\mathrm{S}}(\lambda^{\alpha}\lambda^{\beta}\lambda^{\gamma}/6)}\partial^{2}F(x_{\star})(s_{\alpha},\partial F(x_{\star}))\partial^{2}F(x_{\star})(s_{\beta},s_{\gamma})\cdot \partial F(x_{\star})+ \\ \notag +\,\underbrace{\frac{(\Sigma \epsilon)^{\alpha}(\Sigma \epsilon)^{\beta}(\Sigma \epsilon)^{\gamma}}{6}}_{\mathcal{I}_{\,\mathrm{S}}(\lambda^{\alpha}\lambda^{\beta}\lambda^{\gamma}/6)}\partial^{2}F(x_{\star})(s_{\beta},\partial F(x_{\star}))\partial^{2}F(x_{\star})(s_{\alpha},s_{\gamma})\cdot \partial F(x_{\star})+ \\ \notag +\,\underbrace{\frac{(\Sigma \epsilon)^{\alpha}(\Sigma \epsilon)^{\beta}(\Sigma \epsilon)^{\gamma}}{6}}_{\mathcal{I}_{\,\mathrm{S}}(\lambda^{\alpha}\lambda^{\beta}\lambda^{\gamma}/6)}\partial^{2}F(x_{\star})(s_{\gamma},\partial F(x_{\star}))\partial^{2}F(x_{\star})(s_{\alpha},s_{\beta})\cdot \partial F(x_{\star})-\\ \notag-\,\underbrace{\frac{(\Sigma \epsilon)^{\alpha}(\Sigma \epsilon)^{\beta}(\Sigma \epsilon)^{\gamma}}{6}}_{\mathcal{I}_{\,\mathrm{S}}(\lambda^{\alpha}\lambda^{\beta}\lambda^{\gamma}/6)}\left(D_{\alpha\beta\gamma}^{\delta}\cdot e_{\delta} - 3A_{\alpha\beta}^{\delta}\cdot\partial^{2}F(x_{\star})(s_{\gamma},e_{\delta})\cdot\partial F(x_{\star})\right).
\end{eqnarray}
This, of course, coincides with the image of a general perturbative solution under the action of the $\mathcal{I}_{\,\mathrm{S}}$ morphism.

Thus we have checked by explicit calculation that the general perturbative general solution of equation $F=0$ is realized as a symmetric deformational solution up to the third order.
\section{Perturbative Characteristics}
\subsection{Beta Field} Let $g_{t}$ be a family of regular endomorphisms of variety $X$, parametrized by the parameter $t$, leaving the point $x_{\star}$ fixed. The morphism $E:\mathrm{Spec}(\mathrm{Pert}_{n,k}(\lambda))\rightarrow X$ allows us to transfer this endomorphism to $\mathrm{Spec}(\mathrm{Pert}_{n,k}(\lambda))$, which, in turn, defines an endomorphism of the ring $\mathrm{Pert}_{n,k}(\lambda)$, which we will also denote by $g_{t}$. In other words, the endomorphism of a variety is locally defined by the endomorphism of its chart, which in our case is a $\mathrm{Pert}_{n,k}(\lambda)$-chart. We will denote by $\beta_{g}$ the vector field associated with this family of endomorphisms. This notation comes from quantum field theory. The role of diffeomorphism $g_{t}$ is played by Weyl transformations defined on partition functions, which form the manifold of all quantum field theories \cite{1,4}.

In algebraic terms, the ring endomorphism $g_{t}$ is defined on the generators of $\mathrm{Pert}_{n,k}(\lambda)$ as follows:
\begin{equation}
    \lambda^{\alpha}\mapsto g_{\alpha_{1}}^{1,\alpha}(t)\lambda^{\alpha_{1}}+\,g_{\alpha_{1}\alpha_{2}}^{2,\alpha}(t)\lambda^{\alpha_{1}}\cdot\lambda^{\alpha_{2}}+\,...+\,g^{k,\alpha}_{\alpha_{1}...\alpha_{k}}(t)\lambda^{\alpha_{1}}\cdot...\cdot\lambda^{\alpha_{k}},
\end{equation}
where $g^{i,\alpha}_{\alpha_{1},...,\alpha_{i}}(t)$ - is the $t$-dependent symbols symmetric on the lower indices. Thus, $\beta_{g}$ is the \emph{derivative} of the ring $\mathrm{Pert}_{n,k}(\lambda)$ obtained by taking the derivative of $t$ at the point $t=0$.

The beta field is an important perturbative characteristic of the affine scheme $X$, since a nonzero beta field on $\mathrm{Pert}_{n,k}(\lambda)$ is an obstacle to the existence of a $n$-dimension $g_{t}$-invariant subvariety containing $x_{\star}$.

\subsection{Gamma Structure} The covariant derivative of a vector field at its zero does not depend on the choice of connection on the variety. Thus, we can associate with this field some connection-invariant operator on the tangent space to the zero of this field. This reasoning leads to the existence of another perturbative characteristic, which we will call the $\gamma_{g}$-structure.
\begin{df}
    Let $R$ be a $\mathbb{K}$-algebra and $Y$ be an affine scheme over $\mathbb{K}$, then we call the set of $R$-figures $\mathrm{Hom}(P(Y),R)$. With each $R$-figure $p$ we associate the ideal $\mathrm{Ker}(p)$ in $P(Y)$ of the polynomial function, which are equal to zero on $p$.
\end{df}
\begin{df}
    We will call the $R$-module $\mathrm{Hom}(P(Y),R\otimes\mathbb{K}[\delta]/(\delta^{2}))$ the tangent space to the $R$-figure $p$ on the variety $Y$. Since we work in the category of punctured affine schemes, in the language of rings, here we mean morphisms mapping $\mathrm{Ker}(p)$ to the ideal $(\delta)$.
\end{df}
For mathematicians, these objects are known as functors of points \cite{5}.
\begin{df}
     Let $g_{t}$ be a family of endomorphisms $P(Y)$ preserving the $R$-figure $p$. Then it induces a family of tangent space morphisms $g_{t}^{*}$. We define $\gamma_{g}$ as the tangent space endomorphism given by $\mathrm{d}_{t}g_{t}^{*}|_{t=0}$.
\end{df}

Consider the following example. Let the family $g_{t}$ of morphisms of an affine manifold $X$ induced the family of ring $P(Y)=\mathrm{Pert}_{n,k+1}(\mu)$ morphisms $g_{t}$ defined by the following formula:
\begin{equation}
    \mu^{\alpha}\mapsto \mu^{\alpha}+u^{\alpha}_{\alpha_{1},...,\alpha_{k+1}}(t)\mu^{\alpha_{1}}\cdot...\cdot\mu^{\alpha_{k+1}}.
\end{equation}
Consider now the $\mathrm{Pert}_{n,k}(\lambda)$-figure $\pi$ on $\mathrm{Spec}(\mathrm{Pert}_{n,k+1}(\mu))$ induced by the canonical projection of $\pi: \mathrm{Pert}_{n,k+1}(\mu)\rightarrow\mathrm{Pert}_{n,k}(\lambda)$. It is easy to see that the endomorphisms of $g_{t}$ preserve this figure. Then the set of morphisms $\mathrm{Hom}(\mathrm{Pert}_{n,k+1}(\mu),\,\mathrm{Pert}_{n, k}(\lambda)\otimes\mathbb{K}[\delta]/(\delta^{2}))$ is indeed a $\mathrm{Pert}_{n,k}(\lambda)$-module with the basis whose elements we will call $f_{\alpha}$. Then we get the action of $\gamma_{g}$ on the basis tangent vectors $f_{\alpha}$:
\begin{equation}
    \gamma_{g} (f_{\alpha})=(k+1)\cdot \dot{u}^{\beta}_{\alpha,\alpha_{1}...\alpha_{k}}(0)\lambda^{\alpha_{1}}\cdot...\cdot\lambda^{\alpha_{k}}\cdot f_{\beta}.
\end{equation}
While the $\beta_{g}$-field on $\mathrm{Pert}_{n,k+1}(\mu)$ is given by 
\begin{equation}
    \beta_{g}(\mu^{\alpha})=\dot{u}^{\alpha}_{\alpha_{1}...\alpha_{k+1}}(0)\mu^{\alpha_{1}}\cdot...\cdot\mu^{\alpha_{k+1}}.
\end{equation}

Thus, we obtain the relation between the $\gamma_{g}$-structure and the $\beta_{g}$-vector field, which meets the initial motivation to determine the structure acting on the tangent space to the perturbative solution.
\section{Conclusion \& Discussion}
The purpose of these notes was to explain in detail how the perturbative description of affine variety in the neighborhood of a smooth point is related to the description via multiple deformations. A secondary goal is to explain the geometric nature of the universal connection between two objects studied by physicists in quantum field theory: the anomalous dimension operator and the beta function at first non-trivial order of the perturbation theory. Therefore, one should consider these notes primarily as an explanation of the procedure for constructing perturbative quantum field theories as multiple nilpotent deformations of functors described in \cite{2}.

However, we expect that the explicit formulas connecting $\mathrm{Pert}$- and $\mathrm{Def}$- charts of manifolds that we have described in this paper will be useful in other deformation theories.
\section*{Appendix A}%
\label{sec:appendix A}
\noindent The proof of the theorem 2.1 is given in this appendix.

For $R$ equal to $\mathrm{Pert}_{n,k}(\lambda)$ or $\mathrm{Def}_{n,k}(\epsilon)$, the set $\mathrm{Hom}(\mathrm{Spec}(R),X)$ of morphisms of punctured schemes contains only those maps $E$ that are induced by elements of the set of morphisms of rings $\mathrm{Hom}(P(X),R)$. In turn, ring morphisms are in one-to-one correspondence with the $R$-solutions of the polynomial system $F_{A}=0$ which define $X$. Thus, we can number the morphisms of the schemes by solving $F_{A}=0$ over $R$. 

We will write $E_{p}$, which implies that the corresponding morphism corresponds to a solution $p$ of the system $F_{A}=0$ over the ring $R$. Thus to prove the inclusion of morphisms sets it is enough to show that
\begin{equation}
\mathrm{Pert}_{n,k}(\lambda)\hookrightarrow\mathrm{Def}_{n,k}(\epsilon).
\end{equation}
First, let us show that the rings are indeed embedded.
\begin{st}
    There exists an injective morphism
    \begin{equation}
    \mathcal{I}_{\,\mathrm{S}}:\mathrm{Pert}_{n,k}(\lambda)\hookrightarrow\mathrm{Def}_{n,k}(\epsilon).
    \end{equation}
    It is defined on the generators by the following formula:
\begin{equation}
\mathcal{I}_{\,\mathrm{S}}(\lambda^{\alpha})=\epsilon_{1}^{\alpha}+...+\epsilon_{k}^{\alpha}.
\end{equation}
\end{st}
\begin{pr}
    This morphism comes from the rings map $\mathbb{K}[\lambda^{\alpha}]\rightarrow \mathbb{K}[\epsilon_{i}^{\alpha}]$. Let us make sure that it preserves the denominator ideals. Indeed:
    \begin{equation}
    \mathcal{I}_{\,\mathrm{S}}\left(\lambda^{\alpha_{1}}\cdot...\cdot\lambda^{\alpha_{k+1}}\right)=\prod_{i=1}^{k+1}\mathcal{I}_{\,\mathrm{S}}\left(\lambda^{\alpha_{i}}\right)=\prod_{i=1}^{k+1}\left(\epsilon_{1}^{\alpha_{i}}+...+\epsilon_{k}^{\alpha_{i}}\right),
\end{equation}
since by expanding the last product, we can see that any terms will contain the product of two $\epsilon$ with the same lower indices, which is zero in $\mathrm{Def}_{n,k}(\epsilon)$. Injectivity follows immediately from the linear independence of monomials consisting of different numbers of $\epsilon$.
\end{pr}
Our goal now is to describe the image of the morphism $\mathcal{I}_{\,\mathrm{S}}$. For this purpose we note that the symmetric group $\mathrm{S}_{k}$ acts on $\mathrm{Def}_{n,k}(\epsilon)$ by permutations
\begin{equation}
\label{eq:16}
\sigma\left(\epsilon_{i}^{\alpha}\right)=\epsilon_{\sigma(i)}^{\alpha}.
\end{equation}
This action canonically induces the action of $\mathrm{S}_{k}$ on the  scheme morphisms set $\mathrm{Hom}(\mathrm{Spec}(\mathrm{Def}_{n,k}(\epsilon)), X)$, given by the formula
\begin{equation}
\sigma(E_{p(\epsilon)})=E_{\sigma(p(\epsilon))}.
\end{equation}
Thus, if we prove $\mathrm{Pert}_{n,k}(\lambda)\simeq\mathrm{Inv}_{\mathrm{S}_{k}}(\mathrm{Def}_{n,k}(\epsilon))$, we immediately get the statement of the theorem.
\begin{st}
    The ring $\mathrm{Pert}_{n,k}(\lambda)$ is a subring of symmetric polynomials in the deformation ring $\mathrm{Def}_{n,k}(\epsilon)$:
    \begin{equation}
        \mathrm{Pert}_{n,k}(\lambda)\simeq \mathrm{Im}(\mathcal{I}_{\,\mathrm{S}})=\mathrm{Inv}_{\mathrm{S}_{k}}(\mathrm{Def}_{n,k}(\epsilon)).
    \end{equation}
\end{st}
\begin{pr}
    We will show that any symmetric polynomial is contained in the image $\mathcal{I}_{\,S}$. We note that the ring $\mathrm{Def}_{n,k}(\epsilon)$ has a natural graduation:
    \begin{equation}
    \mathrm{Def}_{n,k}(\epsilon)=\bigoplus_{m=1}^{k}\mathrm{Def}_{n,k}^{(m)}(\epsilon),
    \end{equation}
    where the basis elements in $\mathrm{Def}_{n,k}^{(m)}(\epsilon)$ has the form
    \begin{equation}
        p^{\alpha_{1},...,\alpha_{m}}_{i_{1},...,i_{m}}(\epsilon)=\epsilon_{i_{1}}^{\alpha_{1}}\cdot...\cdot\epsilon_{i_{m}}^{\alpha_{m}}.
    \end{equation}
    Here $\alpha_{l}\in\overline{1,n}$ and $1\leq i_{1}<...<i_{m}\leq k$. Thus an arbitrary polynomial from $\mathrm{Def}_{n,k}^{(m)}(\epsilon)$ can be decomposed by this basis:
    \begin{equation}
    \label{eq:decomp}
        p(\epsilon)=\sum_{1\leq i_{1}<...<i_{m}\leq k} T^{\,i_{1},...,i_{m}}_{\alpha_{1},...,\alpha_{m}}\epsilon_{i_{1}}^{\alpha_{1}}\cdot...\cdot\epsilon_{i_{m}}^{\alpha_{m}}.
    \end{equation}
    Now let $p(\epsilon)$ be such that for any $\sigma\in\mathrm{S}_{k}$ it is true that $\sigma(p(\epsilon))=p(\epsilon)$.
    
    Fix one set $j_{1},...,j_{m}$ such that $1\leq j_{1}<...<j_{m}\leq k$. Consider all permutations that preserve this set. From the invariance of $p(\epsilon)$ with respect to such $\sigma$ it follows that (only the Greek indices are summarized here)
    \begin{equation}
    T^{\,j_{1},...,j_{m}}_{\alpha_{1},...,\alpha_{m}}\epsilon_{\sigma^{-1}(j_{1})}^{\alpha_{1}}\cdot...\cdot\epsilon_{\sigma^{-1}(j_{m})}^{\alpha_{m}}=T^{\,j_{1},...,j_{m}}_{\alpha_{1},...,\alpha_{m}}\epsilon_{j_{1}}^{\alpha_{1}}\cdot...\cdot\epsilon_{j_{m}}^{\alpha_{m}}.
    \end{equation}
    On the other hand, since $\sigma$ leaves the set $j_{1},...,j_{m}$ invariant, the left-hand side effectively reduces to a permutation $\tau\in\mathrm{S}_{m}$ of order of the Greek indices: \begin{equation}
        T^{\,j_{1},...,j_{m}}_{\alpha_{1},...,\alpha_{m}}\epsilon_{\sigma^{-1}(j_{1})}^{\alpha_{1}}\cdot...\cdot\epsilon_{\sigma^{-1}(j_{m})}^{\alpha_{m}}=T^{\,j_{1},...,j_{m}}_{\alpha_{\tau(1)},...,\alpha_{\tau(m)}}\epsilon_{j_{1}}^{\alpha_{1}}\cdot...\cdot\epsilon_{j_{m}}^{\alpha_{m}}.
    \end{equation}
    Moreover, while $\sigma$ runs through the whole $\mathrm{Bij}(\{j_{1},...,j_{m}\})$ the permutation $\tau$ runs through the whole $\mathrm{S}_{m}$. Then it turns out that for any $\tau\in\mathrm{S}_{m}$
    \begin{equation}
    T^{\,j_{1},...,j_{m}}_{\alpha_{\tau(1)},...,\alpha_{\tau(m)}}\epsilon_{j_{1}}^{\alpha_{1}}\cdot...\cdot\epsilon_{j_{m}}^{\alpha_{m}}=T^{\,j_{1},...,j_{m}}_{\alpha_{1},...,\alpha_{m}}\epsilon_{j_{1}}^{\alpha_{1}}\cdot...\cdot\epsilon_{j_{m}}^{\alpha_{m}}.
    \end{equation}
    Thus, in the expansion \eqref{eq:decomp} of the invariant polynomial, all symbols $T_{\alpha_{1},...,\alpha_{m}}^{\,i_{1},...,i_{m}}$ are completely symmetric on the lower indices. Now act by a general permutation $\sigma^{-1}\in\mathrm{S}_{k}$ on the decomposition of the invariant polynomial \eqref{eq:decomp}:
    \begin{equation}
        \sigma^{-1}(p(\epsilon))=\sum_{1\leq i_{1}<...<i_{m}\leq k} T^{\,i_{1},...,i_{m}}_{\alpha_{1},...,\alpha_{m}}\epsilon_{\sigma^{-1}(i_{1})}^{\alpha_{1}}\cdot...\cdot\epsilon_{\sigma^{-1}(i_{m})}^{\alpha_{m}}.
    \end{equation}
    Consider $\sigma^{-1}$ which maps a given ordered set $j_{1},...,j_{m}$ to another ordered set $l_{1},...,l_{m}$ of the same length. From the invariance of $p(\epsilon)$ it follows that \begin{equation}
        T^{\,j_{1},...,j_{m}}_{\alpha_{1},...,\alpha_{m}}\epsilon_{l_{1}}^{\alpha_{1}}\cdot...\cdot\epsilon_{l_{m}}^{\alpha_{m}}=T^{\,l_{1},...,l_{m}}_{\alpha_{1},...,\alpha_{m}}\epsilon_{l_{1}}^{\alpha_{1}}\cdot...\cdot\epsilon_{l_{m}}^{\alpha_{m}}.
    \end{equation}
    Thus, at fixed lower indices, the symbols $T^{\,i_{1},...,i_{m}}_{\alpha_{1},...,\alpha_{m}}$ are equal to each other for any ordered sets of upper indices. Then we get that 
    \begin{eqnarray}
        p(\epsilon)=T^{\,1,...,m}_{\alpha_{1},...,\alpha_{m}}\sum_{1\leq i_{1}<...<i_{m}\leq k} \epsilon_{i_{1}}^{\alpha_{1}}\cdot...\cdot\epsilon_{i_{m}}^{\alpha_{m}}=\\ \notag=\,\frac{1}{m!}\cdot T^{\,1,...,m}_{\alpha_{1},...,\alpha_{m}}\cdot(\epsilon^{\alpha_{1}}_{1}+...+\epsilon^{\alpha_{1}}_{k})\cdot... \cdot(\epsilon^{\alpha_{m}}_{1}+...+\epsilon^{\alpha_{m}}_{k})\in\mathrm{Im}(\mathcal{I}_{\,\mathrm{S}}),
    \end{eqnarray}
    which completes the proof of the theorem.
\end{pr}
\section*{Appendix B}
Morally speaking, the statement of the main theorem is that perturbative solutions of the nonlinear equation $F=0$ are symmetric solutions over the deformation ring. It is technically easier to construct solutions over the deformation ring, and we will now demonstrate this, by computing recurrence formulas expressing $k$-order deformation charts through the previous ones. 

The main advantage is based on the chain of rings inclusions:
\begin{equation}
    \mathrm{Def}_{n,1}(\epsilon)\hookrightarrow\mathrm{Def}_{n,2}(\epsilon)\hookrightarrow...\hookrightarrow\mathrm{Def}_{n,k}(\epsilon)\simeq(\mathrm{Def}_{n,1}(\epsilon))^{\otimes k},
\end{equation}
while for perturbation rings we have
\begin{equation}
    \mathrm{Pert}_{n,1}(\lambda)\twoheadleftarrow\mathrm{Pert}_{n,2}(\lambda)\twoheadleftarrow...\twoheadleftarrow\mathrm{Pert}_{n,k}(\lambda).
\end{equation}
This means that a solution over $\mathrm{Def}_{n,k}(\epsilon)$ is also a solution over $\mathrm{Def}_{n,k+1}(\epsilon)$, while this is not true over a perturbative ring.

Consider a polynomial $p\in\mathrm{Def}_{n,k}(\epsilon)$. It is given by the decomposition:
\begin{equation}
p(\epsilon_{1},...,\epsilon_{k})=p_{0}+p_{1,\alpha}\epsilon_{1}^{\alpha}+\sum_{m=2}^{k}p_{m,\alpha}(\epsilon_{1},...,\epsilon_{m-1})\epsilon^{\alpha}_{m}.
\end{equation}
For any $0\leq i \leq k$ let us introduce the notation $p_{i}(\epsilon_{1},...,\epsilon_{i})$ which are partial sums of this expansion up to the summand of the form $p_{i,\alpha}(\epsilon_{1},...,\epsilon_{i-1})\epsilon_{i}^{\alpha}$. It is clear that such partial sums will depend only on $\epsilon_{1},...,\epsilon_{i}$. 

Let $p(\epsilon_{1},...,\epsilon_{k})$ now be a column whose components $p^{a}(\epsilon_{1},...,\epsilon_{k})$ are polynomials from $\mathrm{Def}_{n,k}(\epsilon)$ such that it solves the equation $F=0$. Then
\begin{equation}
F(p_{0})+\braket{p_{1,\alpha},\partial F(p_{0})\epsilon_{1}^{\alpha}}+\sum_{m=2}^{k}\braket{p_{m,\alpha}(\epsilon_{1},...,\epsilon_{m-1}),\partial F(p_{m-1})}\epsilon^{\alpha}_{m}=0.
\end{equation}
From this equation it immediately follows that $p_{0}=x_{\star}, p_{1,\alpha}\in T_{x_{\star}}X$ and
\begin{equation}
\braket{p_{m,\alpha}(\epsilon_{1},...,\epsilon_{m-1}),\partial F(p_{m-1})}=0.
\end{equation}
Thus, in this sense, we can say that $p_{m,\alpha}(\epsilon_{1},...,\epsilon_{m-1})$ lies in formal tangent space to the point $x_{\star}$. For the sake of brevity, we will write $p_{m,\alpha}\in T_{p_{m-1}}X$.

Thus, knowing the solution in the ring $\mathrm{Def}_{n,k}(\epsilon)$ and some tangent vectors to it, we can construct the solution in $\mathrm{Def}_{n,k+1}(\epsilon)$ simply by deforming the solution by a tangent vector with a new nilpotent element $\epsilon_{k+1}^{\alpha}$. We need to figure out how to construct a tangent space to a solution of order $k$ knowing all the previous ones. 
\begin{theorem}
    Let $p(\epsilon_{1},...,\epsilon_{k})$ be a solution of the equation $F=0$. It can be decomposed in the following form:
\begin{equation}
\label{exp}
p(\epsilon_{1},...,\epsilon_{k})=r(\epsilon_{1},...,\epsilon_{k-1})+t_{\alpha}(\epsilon_{1},...,\epsilon_{k-1})\epsilon_{k}^{\alpha}.
\end{equation}
Let $s, u_{\alpha}\in T_{r}X$. Then general form of $q\in T_{p}X$ giving by the formula:
\begin{equation}
    q=s-\epsilon_{k}^{\alpha}\cdot\sum_{B=1}^{M}\frac{\partial^{2}F(r)(s,t_{\alpha})}{\braket{\partial F(r),\partial F(r)}}\cdot \partial F(r)+\epsilon_{k}^{\alpha}\cdot u_{\alpha}.
\end{equation}
\end{theorem}
\begin{pr}
Let us substitute expansion \eqref{exp} into the system that defines the tangent space. It has the form $\braket{q(\epsilon_{1},...,\epsilon_{k}),\partial F(p(\epsilon_{1},...,\epsilon_{k}))}=0$. Then we obtain
\begin{equation}
\label{eq:35}
\braket{q,\partial F(r)}+\epsilon_{k}^{\alpha}\partial^{2}F(r)(t_{\alpha},q)=0.
\end{equation}
Then the general solution of this equation is
\begin{equation}
\label{eq:36}
q=s-\epsilon_{k}^{\alpha}\cdot\sum_{B=1}^{M}\frac{\partial^{2}F(r)(s,t_{\alpha})}{\braket{\partial F(r),\partial F(r)}}\cdot \partial F(r)+\epsilon_{k}^{\alpha}\cdot u_{\alpha}.
\end{equation}
It is necessary to make a comment that 
    \begin{equation}
     \braket{\partial F(r(\epsilon_{1},...,\epsilon_{k-1})),\partial F(r(\epsilon_{1},...,\epsilon_{k-1}))}\in\mathrm{Def}_{n,k-1}^{*}(\epsilon).
    \end{equation}
    This is clear since for reversibility this polynomial must contain a nonzero free term. The free term of $r$ is exactly $x_{\star}$, i.e. $r^{a}=x_{\star}+\tilde{r}$, and $\tilde{r}$ is nilpotent element. Hence, the free term in $\partial_{d}F(r(\epsilon_{1},...,\epsilon_{k-1}))$ is simply $\partial_{d}F(x_{\star})$, and then by the assumption of regularity of the point $x_{\star}$ the scalar square $\braket{\partial F(x_{\star}),\partial F(x_{\star}})$ is equal to $1$.
\end{pr}
\subsection*{Acknowledgements}
The first author was supported by the Ministry of Science and Higher Education of the Russian Federation (agreement no.075-15-2025-013).

\setcounter{equation}{0}
\renewcommand\theequation{A\arabic{equation}}

\end{document}